\documentclass[manuscript]{aastex}

\shorttitle{A protosolar nebula origin for the ices agglomerated by Comet 67P/Churyumov-Gerasimenko}
\shortauthors{Mousis et al.}

\usepackage{etex}
\usepackage{m-pictex, m-ch-en}
\usepackage {amsmath}
\usepackage[squaren, Gray, cdot]{SIunits}
\usepackage{color}

\begin{document}


\title{A protosolar nebula origin for the ices agglomerated by Comet 67P/Churyumov-Gerasimenko}


\author{O. Mousis\altaffilmark{1}, J. I. Lunine\altaffilmark{2}, A. Luspay-Kuti\altaffilmark{3}, T. Guillot\altaffilmark{4}, B. Marty\altaffilmark{5}, M. Ali-Dib\altaffilmark{6,7}, P. Wurz\altaffilmark{8}, K. Altwegg\altaffilmark{8}, A. Bieler\altaffilmark{9,8}, M. H\"assig\altaffilmark{8,3}, M. Rubin\altaffilmark{8}, P. Vernazza\altaffilmark{1}, J. H. Waite\altaffilmark{3}}


\altaffiltext{1}{Aix Marseille Universit{\'e}, CNRS, LAM (Laboratoire d'Astrophysique de Marseille) UMR 7326, 13388, Marseille, France {\tt olivier.mousis@lam.fr}}
\altaffiltext{2}{Center for Radiophysics and Space Research, Space Sciences Building Cornell University,  Ithaca, NY 14853, USA}
\altaffiltext{3}{Department of Space Research, Southwest Research Institute, 6220 Culebra Rd., San Antonio, TX 78228, USA}
\altaffiltext{4}{Laboratoire J.-L. Lagrange, Universit{\'e} de Nice-Sophia Antipolis, CNRS, Observatoire de la C{\^o}te d'Azur, BP 4229, 06304, Nice, France}
\altaffiltext{5}{CRPG-CNRS, Nancy-Universit\'e, 15 rue Notre Dame des Pauvres, 54501 Vandoeuvre-l\`es-Nancy, France}
\altaffiltext{6}{Universit\'e de Franche-Comt\'e, Institut UTINAM, CNRS/INSU, UMR 6213, Besan\c con Cedex, France}
\altaffiltext{7}{Centre for Planetary Sciences, University of Toronto, Department of Physical \& Environmental Sciences, 1265 Military Trail, Toronto, Ontario, Canada, M1C 1A4}
\altaffiltext{8}{Physikalisches Institut, University of Bern, Sidlerstrasse 5, CH-3012 Bern, Switzerland}
\altaffiltext{9}{Department of Climate and Space Science and Engineering, University of Michigan, 2455 Hayward Street, Ann Arbor, Michigan 48109, USA}


\begin{abstract}
The nature of the icy material accreted by comets during their formation in the outer regions of the protosolar nebula is a major open question in planetary science. Some scenarios of comet formation predict that these bodies agglomerated from crystalline ices condensed in the protosolar nebula. Concurrently, alternative scenarios suggest that comets accreted amorphous ice originating from the interstellar cloud or from the very distant regions of the protosolar nebula. On the basis of existing laboratory and modeling data, we find that the N$_2$/CO and Ar/CO ratios measured in the coma of the Jupiter family comet 67P/Churyumov-Gerasimenko by the ROSINA instrument aboard the European Space Agency's Rosetta spacecraft match those predicted for gases trapped in clathrates. If these measurements are representative of the bulk N$_2$/CO and Ar/CO ratios in 67P/Churyumov-Gerasimenko, it implies that the ices accreted by the comet formed in the nebula and do not originate from the interstellar medium, supporting the idea that the building blocks of outer solar system bodies have been formed from clathrates and possibly from pure crystalline ices. Moreover, because 67P/Churyumov-Gerasimenko is impoverished in Ar and N$_2$, the volatile enrichments observed in Jupiter's atmosphere cannot be explained solely via the accretion of building blocks with similar compositions and require an additional delivery source. A potential source may be the accretion of gas from the nebula that has been progressively enriched in heavy elements due to photoevaporation.

\end{abstract}


\section{Introduction}

Two main reservoirs of ices have taken part concurrently in the formation of icy planetesimals in the protosolar nebula (PSN). The first reservoir, located in a region within 5 to 30 AU of the Sun, contains ices originating from the interstellar medium (ISM), which, due to their near solar vicinity, were initially vaporized when they entered the disk (Chick \& Cassen 1997; Mousis et al. 2000). Later, during the cooling of the PSN, water condensed at $\sim$150 K in the form of microscopic crystalline ice (Kouchi et al. 1994). Planetesimals formed in this region of the PSN are believed to have agglomerated from clathrates, as long as crystalline water was available to form these structures at lower temperatures during the disk's cooling (Mousis et al. 2000, 2012; Iro et al. 2003; Lectez et al. 2015). If water was not sufficiently abundant, or if the kinetics of clathration was too slow, then pure crystalline ices may have formed as well. The other reservoir, located at larger heliocentric distances, is composed of ices that originated from the ISM (Notesco et al. 2003; Bar-Nun et al. 2007) or from the outermost layers of the disk (Ciesla 2014) and that did not vaporize when entering into the PSN (Visser et al. 2011). In this reservoir, the frozen water is amorphous and its porous structure allowed the trapping of other gases (Bar-Nun et al. 1987, 1988, 2007; Bar-Nun \& Laufer 2003). Because the location of the boundary between the two ice reservoirs is poorly determined, it is still unclear from which of the two types of icy materials Oort Cloud comets (OCCs) and Jupiter Family comets (JFCs) preferentially accreted. The uncertainty is reinforced by recent dynamical models suggesting that both OCCs and JFCs could originate from the same parent population, which is the primordial trans-Neptunian disk located in the $\sim$10--40 AU region (Brasser \& Morbidelli 2013).

The recent N$_2$, CO, and Ar measurements (Rubin et al. 2015; Balsiger et al. 2015) performed by ROSINA (Rosetta Orbiter Spectrometer for Ion and Neutral Analysis; Balsiger et al. 2007) aboard the European cometary space mission Rosetta allowed us to derive some clues on the nature of the ices accreted by comet 67P/Churyumov-Gerasimenko (hereafter 67P/C-G) from the comparison of observations with laboratory and modeling data. ROSINA measured the two ratios N$_2$/CO and Ar/N$_2$ over a similar time period in October 2014\footnote{The N$_2$/CO and Ar/N$_2$ ratios have been measured on October 17--23, 2014 (Rubin et al. 2015)  and October 19--23, 2014 (Balsiger et al. 2015), respectively.}. The combination of these two determinations, namely (0.17--1.6) $\times$ 10$^{-2}$ for N$_2$/CO and (9.1 $\pm$ 0.3) $\times$ 10$^{-3}$ for Ar/N$_2$, allowed the inference of  an Ar/CO ratio of (0.15--1.50) $\times$ 10$^{-4}$ in 67P/C-G. In the following, it is assumed that the bulk N$_2$/CO and Ar/CO ratios in 67P/C-G are within the inferred ranges of measurements.

\section{Comparison with laboratory data and models}

Figure 1 shows a comparison between the N$_2$/CO and Ar/CO ratios measured in 67P/C-G, and the same ratios i) sampled from gas trapped in amorphous ice synthesized in the laboratory at temperatures (24, 27, and 30 K) and deposition rates appropriate for the ISM (Bar-Nun et al. 2007), ii) calculated in the case where N$_2$, Ar and CO were crystalized as pure ices in the PSN\footnote{Because CO and N$_2$ are predicted to be the dominating C-- and N-- bearing species in the PSN (Lewis and Prinn 1980), we assume here that CO and N$_2$ are derived from C and N protosolar abundances (Lodders 2009).} and iii) determined in clathrates via the use of a statistical thermodynamic model (van der Waals \& Platteeuw 1959; Lunine \& Stevenson 1985; Sloan \& Koh 2008; Mousis et al. 2010) as a function of their formation temperature (see Appendix for our choice of laboratory data). The amorphous ice has been assumed to be formed in the laboratory from a gas mixture consisting of H$_2$O:CO:N$_2$:Ar = 100:100:14:1, a range of values consistent with those estimated for the protosolar nebula (Bar-Nun et al. 2007). As shown in Figure 1, the N$_2$/CO ratios determined from the experimental measurements of gas trapping in amorphous water ice are within the range obtained from ROSINA measurements for 67P/C-G. Meanwhile, the Ar/CO ratios from these experiments disagree by almost two orders of magnitude with the Ar/CO range for 67P/C-G. 

Moreover, trapping in amorphous ice should not match the Ar/CO ratio at higher deposition temperatures. At the low deposition rate used in these experiments (Bar-Nun et al. 2007), the amount of trapped gas decreases with increasing temperature, because gases can escape easier from the amorphous pores at higher thermal energies (Notesco et al. 1999, 2003). Since the trapping efficiency of Ar and CO in amorphous ice is coincidentally the same (Bar-Nun et al. 1987, 1988, 2007; Bar-Nun \& Laufer 2003), at higher temperatures the Ar/CO ratio should remain approximately similar (e.g. 50 K) to those at $\sim$25 K, despite of a decrease of the total amount of trapped gas in the amorphous pores. Also, CO trapping is 20--70 times more efficient than N$_2$ trapping in amorphous ice, as a result of the weaker bonding of N$_2$ to H$_2$O (Bar-Nun et al. 2007). This difference, along with the faster moving molecules at higher temperatures should lead to a lower N$_2$/CO ratio compared to that at low temperatures. Hence, the amount of N$_2$ relative to CO trapped in amorphous ice should drop below the range deduced for 67P/C-G at  temperatures higher than those used in laboratory measurements.

On the other hand, the ROSINA measurements are not consistent with the possibility that N$_2$, Ar and CO only crystallized as pure ices in the PSN. Because these species condense in the same temperature range in the PSN ($\sim$20--25 K; Mousis et al. 2010), the N$_2$/CO and Ar/CO ratios are predicted to be quasi-protosolar while they are found depleted by factors of at least $\sim$10 and 90 in 67P/C-G's coma, respectively.


\begin{figure}[h]
\begin{center}
\resizebox{\hsize}{!}{\includegraphics[angle=0]{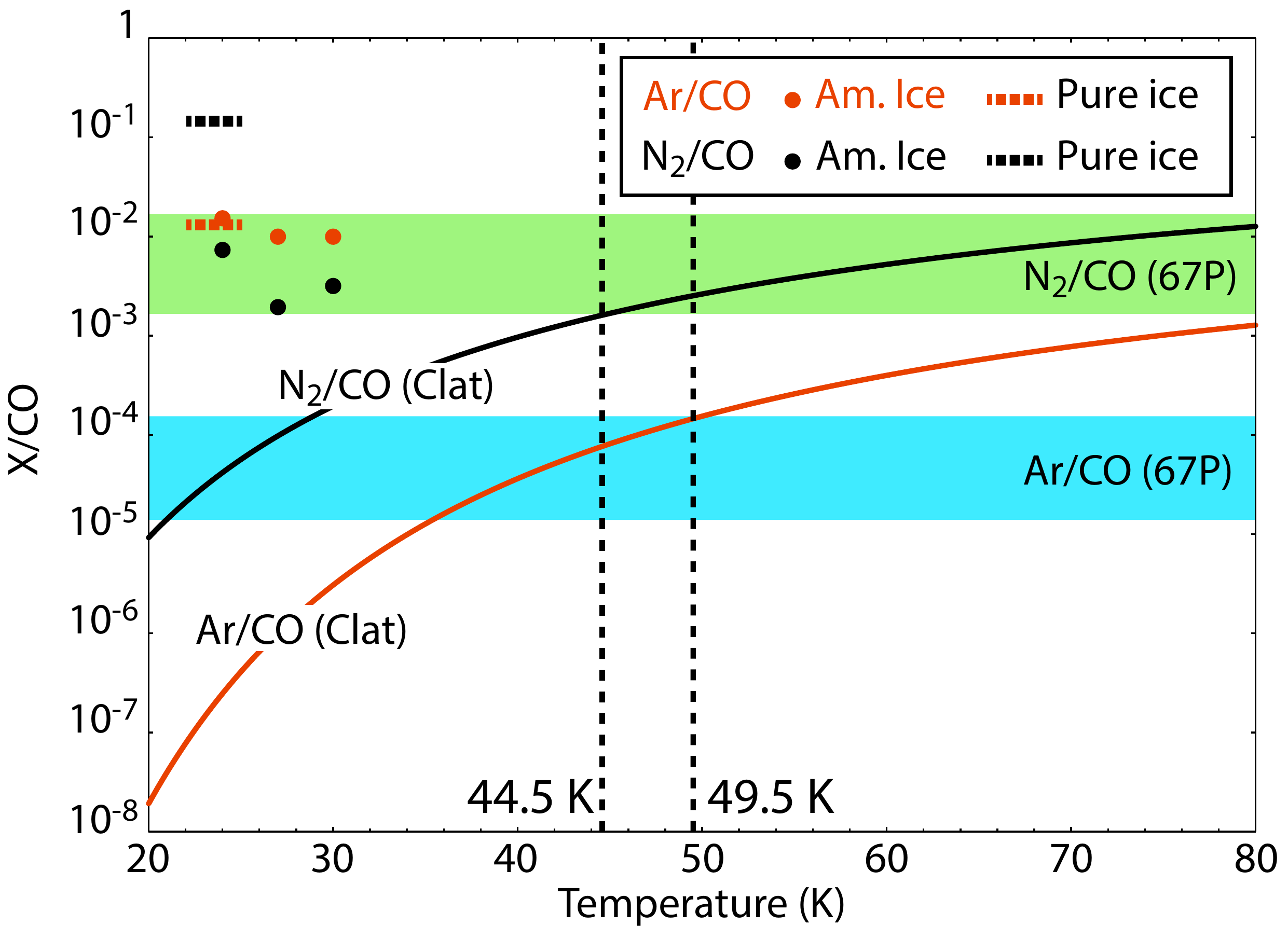}}
\caption{N$_2$/CO and Ar/CO ratios measured by ROSINA in 67P/C-G compared to laboratory data and models. The green and blue areas represent the variation of the N$_2$/CO and Ar/CO ratios measured by ROSINA, respectively (Rubin et al. 2015; Balsiger et al. 2015). The black and red curves represent the evolution of the N$_2$/CO and Ar/CO ratios computed in clathrate as a function of equilibrium temperature, respectively. The black and red dots represent the laboratory measurements of N$_2$/CO and Ar/CO ratios trapped in amorphous ice, respectively (Bar-Nun et al. 2007). The black and red dashed bars represent the N$_2$/CO and Ar/CO ratios at the condensation temperatures of pure crystalline ices. The two vertical dashed lines delimitate the temperature range allowing clathrate formation with N$_2$/CO and Ar/CO ratios consistent with the values measured in 67P/C-G.}
\label{fig:pits}
\end{center}
\end{figure}

Because of the scarcity of experimental data at low-temperature (20--200 K) and low-pressure conditions (10$^{-13}$--10$^{-3}$ bar), which are typical of those encountered in planetary environments, and for the sake of comparison with the ROSINA measurements, we have computed the evolution of clathrate composition along its equilibrium curve at temperatures corresponding to those met during the cooling of the PSN. Here, clathrates are assumed to be formed from water ice grains in contact with a gas mixture composed of CO, N$_2$, and Ar, with CO:N$_2$:Ar = 100:14:1, which is similar to the one used in the amorphous ice experiments. This mixture corresponds to x$_{CO}$~=~0.87, x$_{N_2}$~=~0.12 and x$_{Ar}$~=~0.01 after normalization. In our calculations, we have assumed the formation of Structure I clathrate because CO, the dominant species in our mixture, can only occupy the 5$^{12}$6$^2$ cavities present in this structure under typical PSN conditions (Mohammadi et al. 2005; Zhu et al. 2014). Additionally, N$_2$ and Ar occupy the 5$^{12}$ cavities present in both clathrate Structures I and II. Our calculations have been performed via the use of a statistical mechanics model (van der Waals \& Platteeuw 1959; Lunine \& Stevenson 1985; Sloan \& Koh 2008; Mousis et al. 2010, 2014) that is commonly utilized in the industry and research. The Kihara parameters for the molecule-water interactions employed in this work are listed in Table \ref{kihara} and correspond to fits to existing laboratory experiments.

At a clathrate formation temperature in the $\sim$44--50 K range in the PSN, the N$_2$/CO and Ar/CO ratios simultaneously match the 67P/C-G data. This range is narrower than the one ($\sim$32--70 K) based only on the ROSINA N$_2$/CO determination in 67P/C-G (Lectez et al. 2015).

\begin{table*}[h]
\caption[]{Adopted parameters for the Kihara potential.}
\begin{center}
\begin{tabular}{lccccc}
\hline
\hline
Molecule   	& $\sigma$ (\AA)	& $ \epsilon/k_B$ (K)	& $a$ (\AA)  	& Reference				\\
\hline
CO            	& 3.1515      		& 133.61     			& 0.3976 		& Mohammadi et al. (2005)	\\
N$_2$       	& 3.13512     		& 127.426     			& 0.3526	 	& Sloan \& Koh (2008)		\\
Xe              	& 3.32968     		& 193.708     			& 0.2357 		& Sloan \& Koh (2008)		\\
Ar              	& 2.9434     		& 170.50     			& 0.184 		& Parrish \& Prausnitz (1972)	\\
Kr             		& 2.9739     		& 198.34     			& 0.230 		& Parrish \& Prausnitz (1972)	\\
\hline
\end{tabular}
\end{center}
$\sigma$ is the Lennard-Jones diameter, $\epsilon$ is the depth of the potential well, and $a$ is the radius of the impenetrable core. 
\label{kihara}
\end{table*}

\section{Discussion}

The underlying assumption of our comparison is that the Ar/CO and N$_2$/CO ratios measured by ROSINA in 67P/C-G's coma correspond to the bulk abundances in the nucleus. Additional sources of CO can be neglected since, at the distance of 3 AU corresponding to the epoch of the measurements, the scale lengths for photodissociation or electron impact dissociation are several hundred thousand km while the spacecraft was at $\sim$8 km from the surface at that time period. By adopting the gas phase composition proposed by Bar-Nun et al. (2007), our calculations do not take into account the other molecules that have been detected in 67P/C-G's coma. Provided that these molecules (such as CO$_2$, CH$_4$, etc) remain minor compounds compared to CO in the PSN gas phase, their inclusion in our model would not alter the results. Similar conclusions have been derived by Bar-Nun \& Laufer (2013) in the case of amorphous ice.

Whatever the icy phase from which they derive, Ar, N$_2$ and CO should be released from layers deeper than the shallow subsurface of the nucleus. Indeed, simulations of 67P/C-G's interior show that the clathrate layer would be located at a depth of $\sim$25 m below the surface after 100 years of heating variation due to orbital evolution while the destabilization/sublimation interfaces of the amorphous ice and pure ices are located at minimum depths of $\sim$10 m and 40 m\footnote{This value is given for pure CO crystalline ice.} over the same time period, respectively (Mousis et al. 2015).

\begin{figure}[h]
\begin{center}
\includegraphics[width=12cm,angle=0]{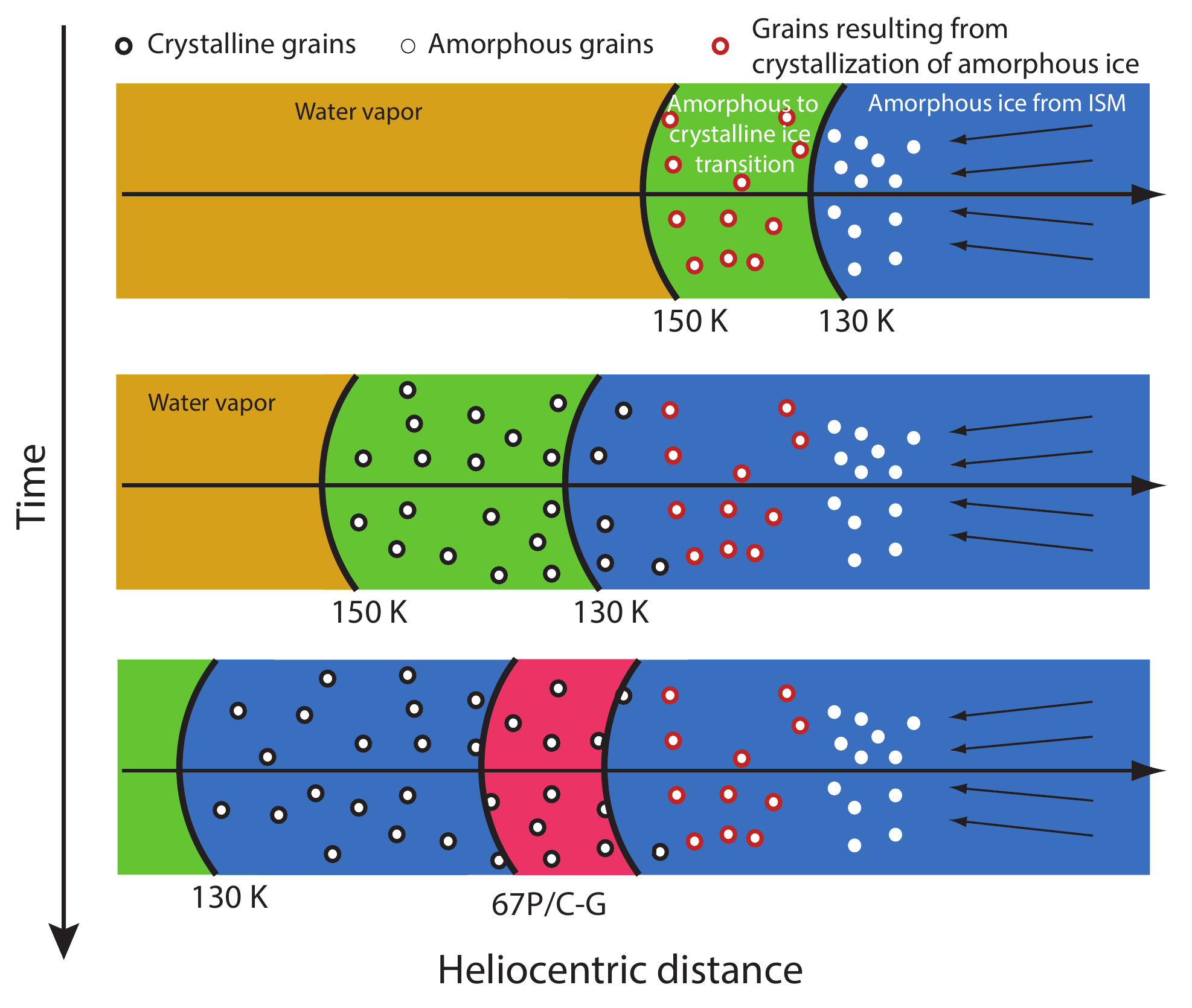}
\caption{The different reservoirs of ice in the PSN. {\it Top panel (early PSN):} the amorphous icy grains (white dots) originating from ISM enter the PSN and migrate inward. In the 130--150 K PSN region, the amorphous ice grains undergo a phase transition and subsequently crystallize (white dots with red circles). At temperatures higher than $\sim$150 K (cf. position of the water snow line), the icy grains delivered to this region vaporize and enrich the gas phase of the disk in H$_2$O vapor. {\it Middle panel (intermediary PSN):} the 130--150 K transition zone has progressed inwards toward the center of the disk. Zones with temperatures lower than 130 K contain crystalline grains (white dots with black circles) that condensed earlier when the disk cooled down to lower than 150 K, crystalline grains from phase transition of amorphous ice (white dots with red circles) but whose structure has changed when the gas phase was in the 130--150 K zone and, at higher heliocentric distances, pristine amorphous grains that never encountered significant warming in the outer PSN. {\it Bottom panel (evolved PSN):} the disk has cooled down to very low temperatures, allowing the crystalline icy grains to adsorb gas molecules from the nebula and form clathrates. The grains agglomerated by 67P/C-G would have formed in this environment (red area).}
\label{fig:depth}
\end{center}
\end{figure}

Based on existing measurements, we show that the N$_2$/CO and Ar/CO ratios observed in 67P/C-G's coma are consistent with the possibility that its nucleus agglomerated from clathrates in the PSN. However, our clathrate composition model simultaneously only fits the highest Ar/CO ratio and the lowest N$_2$/CO ratio observed in 67P/C-G. Because the production rates of Ar and N$_2$ are well correlated in 67P/C-G's coma (Balsiger et al. 2015), a perfect match of our model with the data would require the simultaneous fits of both maximum or minimum N$_2$/CO and Ar/CO ratios. To enable this matching, an additional reservoir of volatiles is needed, which could consist of crystalline ices located deeper than the clathrate layer or amorphous ice located closer to the surface. The additional release of N$_2$ from one of these two reservoirs would then make easier the correlation between the N$_2$/CO and Ar/CO ratios. This points towards a heterogeneity of the type of ices present in the nucleus, unless the Ar/CO and N$_2$/CO ratios observed in the coma are not representative of its bulk composition. So far, only geographical heterogeneity has been discussed (H\"assig et al. 2015; Luspay-Kuti et al. 2015). 

The agglomeration of 67P/C-G from clathrates and pure crystalline ices is consistent with the measurement of its D/H ratio by ROSINA if the comet was assembled from water ice that crystallized at $\sim$140--150 K in the PSN subsequently to the isotopic exchange of D--rich H$_2$O vapor (originating from ISM) with protosolar H$_2$ (Ceccarelli et al. 2014; Kavelaars et al. 2011; Altwegg et al. 2015). In this picture, clathrates formed later at cooler disk temperatures (in the $\sim$40--50 K range as derived from our model) as a result of the interaction of the water ice grains with CO, N$_2$ and Ar vapors. Volatiles not trapped in clathrates then remained in the gas phase until the disk cooled enough to allow their condensation as pure crystalline ices in the $\sim$20--25 K range (Mousis et al. 2010). Because of its smaller propensity to be trapped in clathrate compared to CO and its larger abundance than that of Ar, N$_2$ should be the dominating crystalline ice condensed in the disk. If the PSN dissipated before cooling down to the condensation temperatures of the remaining volatiles, then only clathrates were agglomerated by 67P/C-G. 

The fact that our comparison points towards the presence of clathrates in 67P/C-G makes the assumption of coexisting amorphous ice more difficult to defend because the needed thermodynamic conditions are incompatible with the condensation of crystalline ices at the formation location of the comet in the PSN. Alternatively, some amorphous grains may have migrated within the PSN and crystallized prior to agglomeration by 67P/C-G. However, these grains are expected to be almost devoid of volatiles because of their progressive annealing (Bar-Nun et al. 2007).

If the suggested ice structure of 67P/C-G is typical of that of comets and planetesimals formed in the outer solar system, this suggests that, at their formation location, the disk was at some point warm enough to vaporize the ice grains from the ISM as they entered the PSN (see Figure 2). Our finding supports current scenarios of giant planet and satellite formation (Gautier \& Hersant 2005; Mousis et al. 2009, 2010, 2014), in which the agglomeration of building blocks from clathrates and pure crystalline ices formed in the PSN is needed to explain their current composition. However, if Jupiter accreted from planetesimals with 67P/C-G-like compositions, this poses the problem of interpreting the C, N, and Ar abundances in a consistent manner since these species are all enriched $\sim$3 times compared to protosolar value in Jupiter's atmosphere (Mousis et al. 2012). 

If these species were sourced from 67P/C-G-like icy planetesimals accreted by the forming planet (Gautier et al. 2001; Mousis et al. 2012, 2014), then matching the carbon enrichment would imply that nitrogen is moderately enriched (only $\sim$1.1 times protosolar), due essentially to the accretion of nebula gas including a quasi protosolar N$_2$ abundance, and because the average N$_2$/CO ratio in 67P/C-G is depleted by a factor $\sim$25 compared to protosolar (Rubin et al. 2015). Also, the amount of Ar supplied to Jupiter by these solids would be negligible compared to what is needed to match the observations, resulting in a protosolar Ar abundance in the envelope. This implies that at least CO and Ar must have been delivered to Jupiter via an additional mechanism, which could be the accretion gas that has been progressively enriched in heavy elements due to the preferential photoevaporation of hydrogen and helium from the PSN (Guillot \& Hueso 2006).

So far, if one considers the existing ROSINA and laboratory data, the scenario of 67P/C-G agglomeration from clathrates formed in the PSN appears the best explanation matching the composition of its coma. Future measurements would be needed to ascertain this theory, with notably the determinations of Kr and Xe abundances in the comet, which are predicted to be almost protosolar relative to CO and H$_2$O (see Appendix for details). Unfortunately, due to safety reasons, the spacecraft orbited the comet at much higher distance than envisaged when it was active at perihelion in August 2015, impeding ROSINA's ability to characterize these species. Moreover, the measurements of Kr and Xe will remain challenging after perihelion due to the activity decrease and their expected low abundances.

Further laboratory experiments are needed to constrain the equilibrium curves of clathrates and their kinetics of formation at low pressures. The Kihara parameters used in our statistical mechanics model derive from fits of experimental data available at much higher temperatures than those investigated here. Also, the temperature dependence of the size of cavities is poorly known in clathrates, which may affect their trapping properties. One potential problem might be also the low formation kinetics of clathrates at nebular conditions, but planetesimal collisions, which induce localized heating and enhanced sublimation, may overcome this issue (Lunine \& Stevenson 1985). Moreover, the existing amorphous experiments may not mimic the true formation conditions of cometary grains if they originate from the ISM in terms of vapor deposit rates. However, only the {\it in situ} sampling of a nucleus by a future space mission will provide definitive answer to the question of 67P/C-G's nature and how it formed in the PSN.

\acknowledgements
O.M. acknowledges support from CNES. This work has been partly carried out thanks to the support of the A*MIDEX project (n\textsuperscript{o} ANR-11-IDEX-0001-02) funded by the ``Investissements d'Avenir'' French Government program, managed by the French National Research Agency (ANR).  K.A., P.W., A.B., M.H., A.B., and M.R. gratefully acknowledge the support by the Swiss National Science Foundation.



\appendix

\section{Choice of Laboratory Data}

The laboratory data of volatile trapping in amorphous ice were chosen as published in Bar-Nun et al. (2007) for the following reasons. First, the authors used a very low ice deposition rate in their experiments. Considering the time scale of $\sim$10$^5$--10$^6$ yr for ice grain formation, the low deposition rates considered in this study seem to be more appropriate for interstellar medium and nebula conditions than deposition rates used in previous experiments (Bar-Nun et al. 1987, 1988; Bar-Nun \& Laufer 2003). Second, this group considered a gaseous mixture composed of H$_2$O, CO, N$_2$ and Ar in protosolar proportions (H$_2$O:CO:N$_2$:Ar = 100:100:14:1) prior to amorphous ice formation, which corresponds to a plausible representation of the gas phase composition that may have existed in the protosolar nebula. We did not consider the more recent experimental data describing the composition of amorphous ice formed at 77 K from independent mixtures of X-H$_2$O, with X = Ar, Kr, Xe, N$_2$, and CO (Yokochi et al. 2012), despite the fact that the authors provided extrapolation laws allowing to determine the fraction of noble gas trapped at lower temperatures. These data are irrelevant in our situation because the authors did not simulate gaseous mixtures including together H$_2$O, CO, N$_2$ and Ar that would represent a plausible gas phase composition of the protosolar nebula, as in Bar-Nun et al. (2007). In this case, as illustrated by the data of Bar-Nun et al. (2007), there is a competition between the trapping of the different species in amorphous ice, and the final outcome of this competition is impossible to derive from simple X-H$_2$O mixtures.

\section{Prediction of Kr and Xe abundances in 67P/C-G}

\begin{figure}[tb]
\centering 
\resizebox{\hsize}{!}{\includegraphics[angle=0]{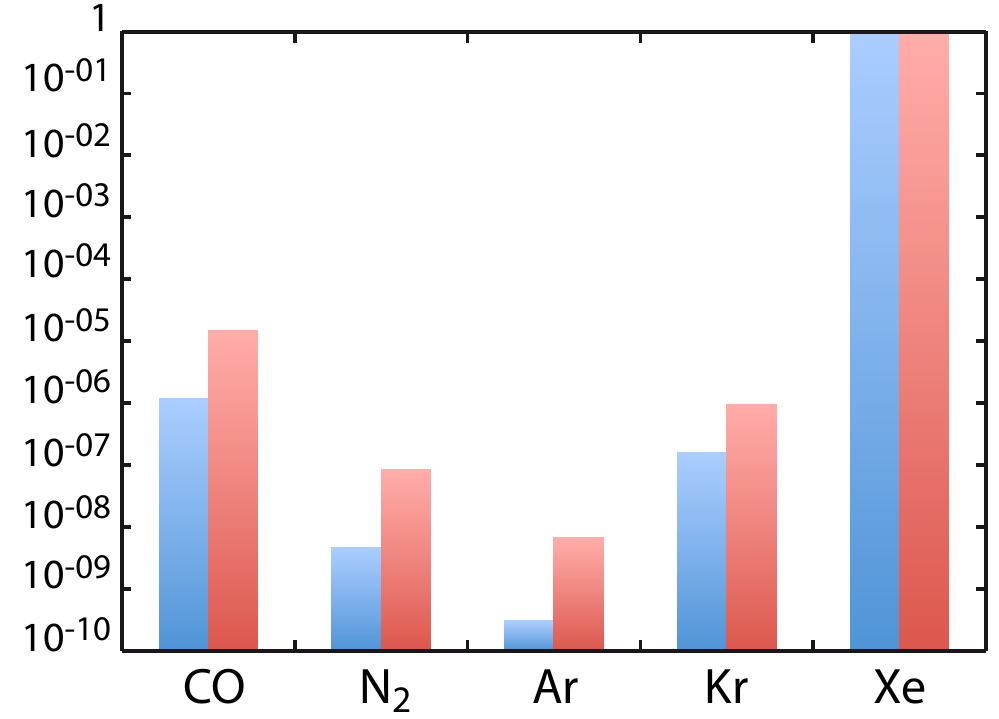}}
\caption[]{Mole fractions of CO, N$_2$, Ar, Kr and Xe in clathrates formed at 44 K (blue bars) and 50 K (red bars) in the protosolar nebula.} \label{fig:KrXe} 
\end{figure}

Our clathrate composition model can also be used to predict the mole fractions of Kr and Xe trapped in 67P/C-G at its formation epoch. To do so, we considered the same gaseous mixture as in Sec. 1 to which we added Kr and Xe, assuming that Kr/Ar and Xe/Ar are in protosolar proportions (Lodders et al. 2009). After normalization, we get x$_{CO}$ = 0.8696, x$_{N_2}$ = 0.1217, x$_{Ar}$ = 0.0087, x$_{Kr}$ = 6.16 $\times$ 10$^{-6}$ and x$_{Xe}$ = 6.02 $\times$ 10$^{-7}$. Figure \ref{fig:KrXe} represents the mole fractions of CO, N$_2$, Ar, Kr and Xe in clathrates formed at 44 and 50 K in the protosolar nebula. Because Xe dominates the clathrate composition, this noble gas should be fully trapped in 67P/C-G, implying Xe/O and Xe/C ratios close to protosolar in the nucleus. Interestingly, the Kr/C ratio in the considered clathrates is in the 0.06-0.14 range, corresponding to enrichment factors that are 3 to 4 magnitude orders larger than the coexisting Kr/C gas phase ratio. This also implies that Kr should be fully trapped in 67P/C-G, thus giving Kr/C and Kr/O ratios close to protosolar in the nucleus.

\end{document}